\documentstyle[aaspp4]{article}

\begin{document}

\title{FUSE Observations of RX~J0513.9$-$6951}

\author{J.B.~Hutchings\altaffilmark{1}, 
A.P.~Cowley\altaffilmark{2}, 
R.~Mann\altaffilmark{1},
P.C.~Schmidtke\altaffilmark{2}, 
\& D.~Crampton\altaffilmark{1} }

\affil{$^1$Herzberg Institute of Astrophysics, NRC of Canada, Victoria, B.C.
V8X 4M6, Canada; ~john.hutchings@nrc.ca}

\affil{$^2$Department of Physics \& Astronomy,
Arizona State University, Tempe, AZ, 85287-1504; anne.cowley@asu.edu, 
paul.schmidtke@asu.edu } 

\begin{abstract}

FUSE observations were obtained in July 2003 during 1.2 cycles of the
0.76-day binary orbit of RX~J0513.9$-$6951.  Radial velocity measurements
of the broad O~VI emission profile show a semiamplitude of K$\sim$26 km
s$^{-1}$, which is much smaller than the value of 117 km s$^{-1}$ measured
from 2001 FUSE data.  Narrow O~VI emissions show no measurable velocity
variation.  The mean velocity of the broad O~VI emission is red-shifted by
$\sim$500 km s$^{-1}$ with respect to both the systemic and narrow
emission-line velocities.  Spectral difference plots show phase-related
changes in the broad emission profile.  Other phase-related changes such
as line and continuum variations are also smaller than in the 2001
spectra.  We describe a moving broad absorption feature near 1020\AA\ as
possible O~VI outflow associated with a precessing jet.  We discuss the
implications for the stellar masses if the 2003 broad O~VI velocities
outline the compact star's orbital motion. 

\end{abstract}


\section{Introduction}

Supersoft X-ray sources are close binary systems thought to contain a
rapidly accreting white dwarf and a low-mass donor star.  Direct mass
determinations for these systems have been difficult to make since the
low-mass star is much fainter than the bright accretion disk.  It seemed
possible that in the Large Magellanic Cloud system RX~J0513$-$695 the far
ultraviolet O~VI emission lines might outline the accreting star's orbital
motion, hence revealing information about the stellar masses. 

RX~J0513$-$6951 is the most luminous of the known supersoft X-ray binary
systems in the Milky Way or the Magellanic Clouds.  The 0.76-day binary
period is well determined from optical light curves observed over many
years by the MACHO project (e.g. Alcock et al.\ 1996; Cowley et al.\ 2002).
The small variation in the orbital light curve ($\Delta$m$\sim$0.06 mag)
and the low-amplitude of the velocity curve from optical emission lines
(semiamplitude K$\sim$11 km s$^{-1}$) suggest the system is viewed nearly
pole-on.  Optical and X-ray fluxes show antiphased long-term high/low
states (Reinsch et al.\ 1996), and optical spectra show the velocity curve
changes between the high and low states (Cowley et al.\ 2002).
Additionally, the system shows evidence of bi-polar jets in some optical
emission lines, with velocities of $\pm$$\sim$4000 km s$^{-1}$ (Crampton
et al.\ 1996; Southwell et al.\ 1996).  There is also evidence of a long-term
photometric period of $\sim$83.2 days (Cowley et al.\ 2002) which might be
related to precession of the accretion disk. 

Using FUSE data obtained in 2001, Hutchings et al.\ (2002) found the O~VI
resonance doublet was composed of a broad, blended emission feature, with
narrow O~VI emission superimposed.  The broad blend showed a moderately
high velocity semi-amplitude (K$\sim$117 km s$^{-1}$), which implied
rather large stellar masses if due to orbital motion of the compact star
(Hutchings et al.\ 2002).  To determine if this motion was a transient
event or a permanent feature of the system, we obtained additional FUSE
spectra in July of 2003. 

\section{2003 Data and Measurements}

RX~J0513.9$-$6951 was observed continuously (with earth occultations) on
2003 July 5 for 1.2 binary orbits.  We use phases based on optical minimum
light from Cowley et al.\ (2002):

T(min) = JD 2,448,858.099$\pm0.001$ + 0.7629434$\pm0.0000020$E days
                 ~~~(= MJD 48857.599)

\noindent
However, we note that although the photometry provides an accurate clock,
the system's low inclination suggests $\Phi$=0 may not occur exactly at
the superior conjunction of the compact star.  Hence, although we know the
relative phasing of variable features, the orientation of the stars with
respect to the observer is somewhat uncertain. 

The FUSE observations were processed as 15 individual spectra through the
binary orbit phases, corresponding to the visibility windows during the
observation period.  Table 1 lists both the total observing time and that
obtained during spacecraft nighttime.  Count rates in all channels were
steady, indicating that the telescope alignments were good.  The long
exposure windows show that we were close to the continuous viewing zone
(CVZ).  The airglow contamination was not correlated with the fraction of
orbital night in each spectrum, so we did not separate the night and day
portions of the observations.  We note that the strength of the airglow
emission in both epochs is similar.  The mean FUSE spectrum from 915\AA\
to 1100\AA\ is shown in Fig.\ 1.  To aid in understanding the spectrum,
the upper dotted line in each panel shows the H$_2$ absorption spectrum 
and the lower dashed line shows the position of the principal airglow 
lines.

\subsection{The Line Spectrum} 

The O~VI doublet (1032, 1038\AA) remains the principal emission feature in
the far ultraviolet spectrum.  O~VI continues to show both a broad (some
20\AA\ wide), blended emission and sharp individual peaks (see Fig.\ 1 \&
2).  He~II (1084\AA) emission is also present, although it is badly
blended with airglow lines, making its measurement very difficult.  In
Fig.\ 1 one can see that He~II has a P-Cygni structure, with absorption on
the violet side of the emission, although contaminated by both H$_2$
absorption and airglow emission.  We looked for, but did not find clear
evidence for other possible emission lines such as N~III (991\AA) and
C~III (977, 1175\AA).  These wavelengths are too close to strong airglow
lines which have been edited out of the spectrum shown in Fig.\ 1. Thus,
emission from either N~III or C~III may contribute to the airglow feature
which was removed.  However, we do see narrow absorptions which are likely
to be C~III and N~III, similar to the He~II absorption, and suggest these
lines may have a P Cyg profile.  We note that optical spectra show weak,
narrow emissions of O~VI, N~V, and C~IV, but lower ionization lines (N~III
and C~III) are probably not present.  

In both the 2001 and 2003 FUSE spectra, absorption lines are found in the
higher Lyman lines (as marked in Fig.\ 1 of Hutchings et al.\ 2002).
Similar features at the Balmer lines are seen in optical spectra (e.g.
Crampton et al.\ 1996). 

In Fig.\ 1 \& 2 we have plotted the H$_2$ absorption spectrum to show the
reader that a number of the absorptions in the spectrum are due to local
molecular hydrogen.  In addition, the interstellar absorption of the C~II
doublet at 1036-7\AA\ is clearly present, with radial velocity $\sim$+30
km s$^{-1}$, as compared to the systemic velocity of +280 km s$^{-1}$ 
(Crampton et al.\ 1996). 

There is a strong dip in the spectra near 1020\AA\ at both epochs.  It
appears to shift to a more positive wavelength in the 2003 spectra.  It
might possibly be associated with the narrow O~VI line at 1032\AA.  If so,
it could be a signature of outflowing gas with velocity near $-$3300 km
s$^{-1}$ and full width about 700 km s$^{-1}$.  A corresponding
absorption associated with the 1038\AA\ O~VI line would be totally 
obscurred by L$\beta$ absorption and by strong airglow emission. 

\subsubsection{ O~VI Emission Lines}

Figure 2 compares the 2001 and 2003 spectra in the region around the broad
O~VI emission feature.  We show the mean spectra from each epoch, with the
continuum flux difference removed by an offset of 1.3e-14 (raising the
possibility of zero-point errors, which we discuss below).  This figure
shows that both the broad and narrow O~VI emissions were stronger during
the 2003 observations. 

The sharp O~VI emission peaks show a mean velocity of +320 km s$^{-1}$,
with a P-Cygni type absorption at $\sim$+30 km s$^{-1}$.  These lines show
a total radial velocity range of only $\sim$15 km s$^{-1}$ and no clear
phase dependence.  This small variation is comparable with measures of the
interstellar absorptions of C~II and H$_2$.  Thus, the scatter may
represent the errors due to the wavelength calibrations and overall S/N.
Note that optical spectra of He~II show a semiamplitude of only K=11 km
s$^{-1}$, which is within our scatter of our FUSE measurements.  The He~II
1084\AA\ line is too badly blended with airglow emission to measure its
velocity. 

Figure 3 shows the broad O~VI mean profile, with the absorptions and
narrow emissions edited out (by removing the sharp airglow emissions and
sharp interstellar absorptions to leave a smooth broad residual).  The
figure also shows the reflected outer wings, centered on the wavelength
that gives the most compact symmetrical profile.  To measure the velocity
of this broad emission we used the mean edited spectrum as a template.
This template was cross-correlated against the unedited individual
spectra, and the centroid of the broad cross-correlation peak was
recorded.  This process was repeated for different versions of the edited
template, and also using variously edited individual spectra.  The
centroids of the broad cross-correlation peaks were not very sensitive to
these detailed variations of process, suggesting that we are indeed
measuring a property of the broad emission.  We also measured the same
velocity shifts by fitting curves through the whole broad feature.  This
too, yielded similar values, but with considerably more scatter. 

The resulting broad-emission O~VI velocity curve from the cross-correlation 
process, described above, is shown in Figure 4.  We note that we have small 
binary phase overlap in our data, and in all the quantities the overlap 
measures are in good agreement, supporting the assumption that we are 
measuring phase-related changes.  Of course, we would prefer to have a 
second full cycle of observations to confirm this.  These cross-correlation 
velocity values were used to fit a sine-curve, whose semiamplitude was found 
to be K$\sim$26 km s$^{-1}$, much lower than the value of K$\sim$117 km 
s$^{-1}$ found in the 2001 data using the same measuring techniques.  
However, the full velocity range is more than twice as large of that of
the optical He~II lines.  The phase of O~VI maximum velocity occurs at 
$\Phi_{phot}$=0.77, compared to $\Phi_{phot}$=0.70 in 2001.  This phase 
difference is formally significant at the 3$\sigma$ level.  While this 
amplitude is less than 1\% of the full width of the feature, it is derived 
from a number of independent spectra, and the values appear to be robust 
to details of the measuring procedure.  We note also that the fit is based 
on measures that cover only 1.2 binary orbits, so that we may be seeing 
changes that are not binary-phase dependent.  Against that, we also note 
that the values in the overlap phases do agree well. 

The mean wavelength of the broad O~VI line corresponds to a velocity of
about +500 km s$^{-1}$ more positive than the mean of the narrow lines or
the known systemic velocity of +280 km s$^{-1}$ (Crampton et al.\ 1996).  
Possible implications of this are discussed below. 

The O~VI broad emission flux was measured by integrating over the entire
feature.  The fluxes from the SiC channels were systematically lower than
the LiF channels, by about 10\%, but the phase-variation was similar in
all channels.  A plot of the mean of the LiF channels and the best-fit sine
curve through them is shown in Fig.\ 4.  The fit is significantly different
from no variation at the 3$\sigma$ level.  The variation range is about
6\%, again much smaller than the $\sim$30\% seen in 2001, and the phasing
(if the variation is real) differs by $\sim$0.2P. 

Figure 5 shows difference plots between the individual spectra and the
mean in the O~VI region.  The differences are heavily smoothed and
amplified several times to reveal the broad changes in the profile, which
appear principally as a dip that moves back and forth.  The diagram also
shows a dotted line indicating how the broad difference dip moves in
orbital phase with a large velocity amplitude.  This was quantified in the
following way.  The dip centroid was measured in each of the difference
plots (roughly in the wavelength range 1025 to 1050\AA), and these values
were used to derive a best-fit sine curve.  The centroids are marked with
dots in Figure 5.  The fit to these with orbital phase has a semi-amplitude
of 3.8$\pm$0.9\AA\ (or K$\sim$1100 km s$^{-1}$), with a zero point value
of 1036.9$\pm$0.6\AA, which is close to the mean wavelength of the O~VI
emission blend.  This too is an indication that we are measuring changes
in the broad emission feature, and not artifacts of the narrower blended
features. Table 2 summarizes the fitted quantities.

\subsubsection{He~II Emission at 1084\AA\ }

In the 2001 data, He~II (1084\AA) was present, although it was blended
with airglow features (the strongest being within 0.1\AA\ of the systemic
velocity of the binary), making it difficult to estimate its strength and
any variations.  The 2003 data have much longer observation windows which
are close to being in the CVZ.  We find that airglow is present at all
times, and cannot be reduced by isolating the night-time data.  He~II
appears to be weaker than in the 2001 spectra, but the width of the
emission ($\sim$4\AA) and its central wavelength make it clear that there
is He~II emission blended with the airglow line.  The He~II line also
shows P-Cygni type absorption on its violet edge (as do the narrow O~VI
emissions).  Unfortunately the combination of airglow and H$_2$ near
1084\AA\ make measurement of He~II impossible in the new FUSE spectra. 

In optical spectra He~II (4686\AA) is always prominent, with wide wings
(comparable with O~VI) and a profile which varies over timescales much
longer than the orbital period (see Crampton et al.\ 1996; Southwell et
al.\ 1996).  In addition, optical spectra also show narrow emission from
the He~II Pickering series lines. 

\subsubsection{Hydrogen Absorption Lines}

The higher Lyman series lines show deep narrow absorptions, as in the 2001
data.  Figure 6 shows some of the lines and the mean, plotted in radial
velocity units.  The absorptions are saturated and have a flat saturated
minimum to very high series numbers, so that we cannot tell if they have
more than one component.  The stronger Lyman lines have sharp (airglow)
emission close to rest velocity, but the figure is made from lines that
show no sign of these.  We measured individual line wavelengths and also
cross-correlated the region from 920-945\AA\ with the mean spectrum.  These
measures indicate a very small phase-related radial velocity variation (K
$\sim$8 km s$^{-1}$), as summarized in Table 2 and shown in Figure 4.
This variation is shifted by $\sim$0.4 in phase from the broad O~VI
emission velocity curve.  It too has much lower amplitude than that found
in the 2001 data, where the semiamplitude was found to be K$\sim$54 km
s$^{-1}$ and the velocity curve was antiphased with the broad O~VI
emission. 

\subsection{The Continuum} 

The 2003 continuum flux is lower by $\sim$30\% (within the measuring
uncertainty of $\sim$10\%) than 2001.  It also shows smaller variations
around the binary orbit.  However, the flux changes from 2001 across the
entire FUSE range are similar (to within 10\%), so there appears to be no 
significant change in the spectral energy distribution with flux level.  
We have no ground-based or FUSE FES photometry to see if the lower flux 
in 2003 corresponded to a fainter optical magnitude. 

Both the continuum flux level and the overall broad emission flux were
measured as a function of orbital phase.  The 2003 spectra show small
($<$10\%) changes in the continuum with phase (see Figure 3), but we do
not observe the 60\% flux rise between photometric phases 0.05 and 0.40
which was seen in the 2001 data.  However, the FUSE light maximum still
occurs near photometric phase $\sim$0.2, as in 2001. 

\subsection{Long Period Variations}
 
Cowley et al.\ (2002) found a long photometric period (83.2 day) in the
MACHO data.  The folded light curve shows an amplitude of $\sim$0.05-0.06
magnitudes at optical wavelengths.  This amplitude is consistent with the
variation being due to inclination changes of a nearly face-on disk
precessing over a small angle.  For this simple assumption, the precession
angle would be $\sim$10$^{\circ}$ or less for small orbital inclinations
($i$ less than 30$^{\circ}$). 

The same scenario would predict changes in the observed jet velocities,
assuming they precesses normal to the accretion disk.  Crampton et al.\
(1996) have measured jet velocities for three epochs, corresponding to
83-day phases of 0.99, 0.34, and 0.35.  The precession scenario fixes the
maximum velocity at maximum light.  Fitting the Crampton et al.\ data to
this model gives a semiamplitude of the jet velocity of K$\sim$160 km
s$^{-1}$, with a mean value near 4000 km s$^{-1}$.  This also gives
possible values of the orbital inclination and precession angle which are
similar to those implied by the low orbital light curve and optical
velocity amplitude. 

Thus, the long-period light curve and the jet velocity changes are
compatible with a precession scenario in which $i$ is less than
30$^{\circ}$ and the precession angle is less than 10$^{\circ}$.  These
values also turn out to be consistent with the mass considerations shown
in Fig.\ 7.  Of course, many more complex scenarios are possible, since the
observational evidence is minimal so far.  Further observations would be
very useful. 

The FUSE observations have (by chance) similar long-period phasing to the
Crampton et al.\ jet measurements.  The 2001 data were taken at phase 0.99
and the 2003 data at phase 0.30.  Although we see no evidence for narrow
jet emission lines in the FUSE data as are found in optical data at  He~II
4686\AA\ and  H$\beta$, the width of the broad O~VI blend is comparable to
the jet velocities.  We also note that the 1020\AA\ absorption could be
outflowing O~VI gas, with a velocity of $-$3100 km s$^{-1}$ ($-$3400 km 
s$^{-1}$ in 2001).  These velocities are lower than the measured jet 
velocities at the same phases, but the change is very similar.  This would 
also be consistent with the precession scenario if the absorption arises 
in an outflow that has a wider opening angle than the jet itself. 

\section{Discussion and Conclusions}

We find the O~VI profile was stronger in 2003 in the innermost $\pm$1800
km s$^{-1}$, but otherwise it still shows a very broad blended profile
with both lines of the doublet having approximately equal strength.  Thus,
the lines presumably arise in an optically thick environment.  The 2003
broad O~VI velocity curve has a small semiamplitude of K$\sim$26 km
s$^{-1}$ (much lower than the 117 km s$^{-1}$ seen in 2001), indicating
that the earlier (and perhaps both) variations are not dominated by
orbital motions.  It is interesting that the derived binary phasing of the
broad velocity variations is close to the same as the 2001 values.  The
maximum velocity occurs near quadrature (assuming minimum light is when
the unheated side of the donor star is most visible).  This is as expected
if the velocities are related to the white dwarf's orbital motion, arising
in its accretion disk.  The change in amplitude, but little in phase, may
indicate radial rather than azimuthal changes in the O~VI disk structure. 

The spectral difference plots show that there are phase-related profile
changes arising over the whole broad profile.  These can be represented as
a wide absorption feature with velocity semiamplitude of K$\sim$1100
kms$^{-1}$.  This is far too high to be orbital motion, but it may arise in
an asymmetry in the hot inner accretion disk around the compact star or be
related to the inner parts of the high velocity jet which presumably lies
near the line of sight.  In this connection we note the broad absorption
dip at $\sim$1020\AA\ might also arise from O~VI moving near the terminal
velocity of the jet. 

Although the O~VI velocity phasing is close to the 2001 value, the line
and continuum flux changes have different phasing.  The narrow O~VI
emission peaks show no significant velocity changes.  The Lyman
absorptions, continuum flux, and O~VI emission flux all have lower
amplitudes than in the 2001 data.  In addition, they all show different
phasing, while they were nearly in phase in the 2001 data.  The Lyman
absorptions are saturated.  They have a mean velocity of
+200 km s$^{-1}$.  No Lyman absorption lies at the sharp O~VI velocity of
+350 km s$^{-1}$, even if there are two components.  The Lyman lines are
variably contaminated by airglow and H$_2$ absorption, so phase-related
profile changes cannot be detected easily.  The small phase-related
velocity change could possibly be due to outflow from the system, as well
as intervening absorbers not related to the binary system.  The phasing of
these velocity variations is consistent with origin in the gas stream from
the donor star. 

From the 2001 data we postulated that the broad O~VI emission might arise
in an orbit close to the white dwarf, and thus may represent its orbital
velocity.  Clearly, the substantial change between observing epochs
indicates that the large velocity amplitude measured in 2001 must have
arisen in mass flows within the system, not just from the white dwarf's
orbital motion.  However, for interest, in Figure 7 we show the implied
masses for a star with the velocity semiamplitude of K=26 km s$^{-1}$
(from the 2003 O~VI broad emission).  The diagram also shows the boundaries
set by assuming the jet velocities of $\pm$4000 km s$^{-1}$ are at the
escape velocity of the white dwarf, and the broad line width ($\sim$2500
km s$^{-1}$) represents the rotational velocity of disk gas near the white
dwarf's surface.  These contraints define an area in the mass plot of
Figure 7 bounded by the dashed lines.  If we further assume that the mass
losing star has evolved off the main sequence, the allowed mass values are
$\sim$0.6-1.2M${\odot}$ for the white dwarf and $\sim$0.3-0.4M${\odot}$
for the donor star. 

Determining the mean velocity of the broad emission involves a number of
assumptions.  If it is symmetrical, as shown in Fig.\ 2, we find it is
redshifted from the sharp emission components by $\sim$+500 km s$^{-1}$,
suggesting a possible gravitational redshift of gas near the surface of
the white dwarf.  The gravitational redshift of a 0.65M$_{\odot}$ WD is
about 300 km s$^{-1}$ and that of a 1.4M$_{\odot}$ WD is 600 km s$^{-1}$.
Thus, the observed shift might be evidence that the WD is massive and that
the O~VI arises very close to its surface.  The fact that the broad doublet
lines are apparently of equal flux, indicates that they are formed in a
high density region. 

We thank Alex Fullerton for help with the FUSE data, and the referee
for helpful comments.

\clearpage

\begin{deluxetable}{ccccc}
\tablehead{\colhead{Dataset} &\colhead{JD (mid-exp)} 
&\multicolumn{2}{c}{Exposure time, in sec}
&\colhead{Photometric} \\
\colhead{D006010....} & 
\colhead{2452800 +} & 
\colhead{Total} &
\colhead{Nighttime} &
\colhead{Phase\tablenotemark{a} }
}
\startdata
1001 &25.81250 &4075 &915  &0.535 \cr
1002 &25.85938 &4010 &980  &0.560 \cr
1003 &25.91016 &3885 &1780 &0.663 \cr
1004 &25.96094 &3936 &1574 &0.729 \cr
1005 &26.00781 &4221 &354  &0.791 \cr
1006 &26.05859 &4256 &1950 &0.857 \cr
1007 &26.11719 &5033 &2110 &0.934 \cr
2001 &26.19922 &4234 &1700 &0.041 \cr
2002 &26.27344 &4190 &1450 &0.139 \cr
2003 &26.34375 &4076 &1110 &0.231 \cr
2004 &26.41797 &4002 &790 &0.328 \cr
2005 &26.49219 &4088 &530 &0.425 \cr
2006 &26.56250 &4280 &400 &0.518 \cr
2007 &26.63672 &4439 &491 &0.615 \cr
2008 &26.70703 &4095 &704 &0.707 \cr

\enddata

\tablenotetext{a}{Phases with respect to light minimum, which is not
necessarily conjunction of the two stars.  Ephemeris: T(min) = JD
2,448,858.099$\pm0.001$ + 0.7629434$\pm0.0000020$E  (= MJD 48857.599),
where T(min) is minimum $V$ light (Cowley et al.\ 2002). } 

\end{deluxetable}

\begin{deluxetable}{cccc}
\tablehead{\colhead{Feature} &\colhead{Photometric} &\colhead{Semiamplitude}
&\colhead{ME} \\
\colhead{} &
\colhead{Phase\tablenotemark{a}} &
\colhead{(km s$^{-1}$ or \%) } &
\colhead{(km s$^{-1}$ or \%) }  
}
\startdata
O~VI broad emis  & 0.77$\pm$0.02 & 26$\pm$4 km s$^{-1}$ & 16 km s$^{-1}$ \cr
O~VI broad diffs & 0.76$\pm$0.04 & 1100$\pm$180 km s$^{-1}$ & 300 km s$^{-1}$ \cr
Ly narrow abs    & 0.40$\pm$0.04 & 8$\pm$2 km s$^{-1}$  &8 km s$^{-1}$ \cr
1070\AA\ continuum flux & 0.16$\pm$0.10 & 1.5$\pm$1.0\% & 2\%\cr
O~VI flux        & 0.93$\pm$0.05 & 2.8$\pm$0.8\%        & 2.0\%\cr 

\enddata
\tablenotetext{a}{Phases with respect to $V$ light minimum. 
Ephemeris:
T(min) = JD 2,448,858.099$\pm0.001$ + 0.7629434$\pm0.0000020$E 
 (from Cowley et al.\ 2002). }

\end{deluxetable}

\clearpage

\begin{figure}
\plotone{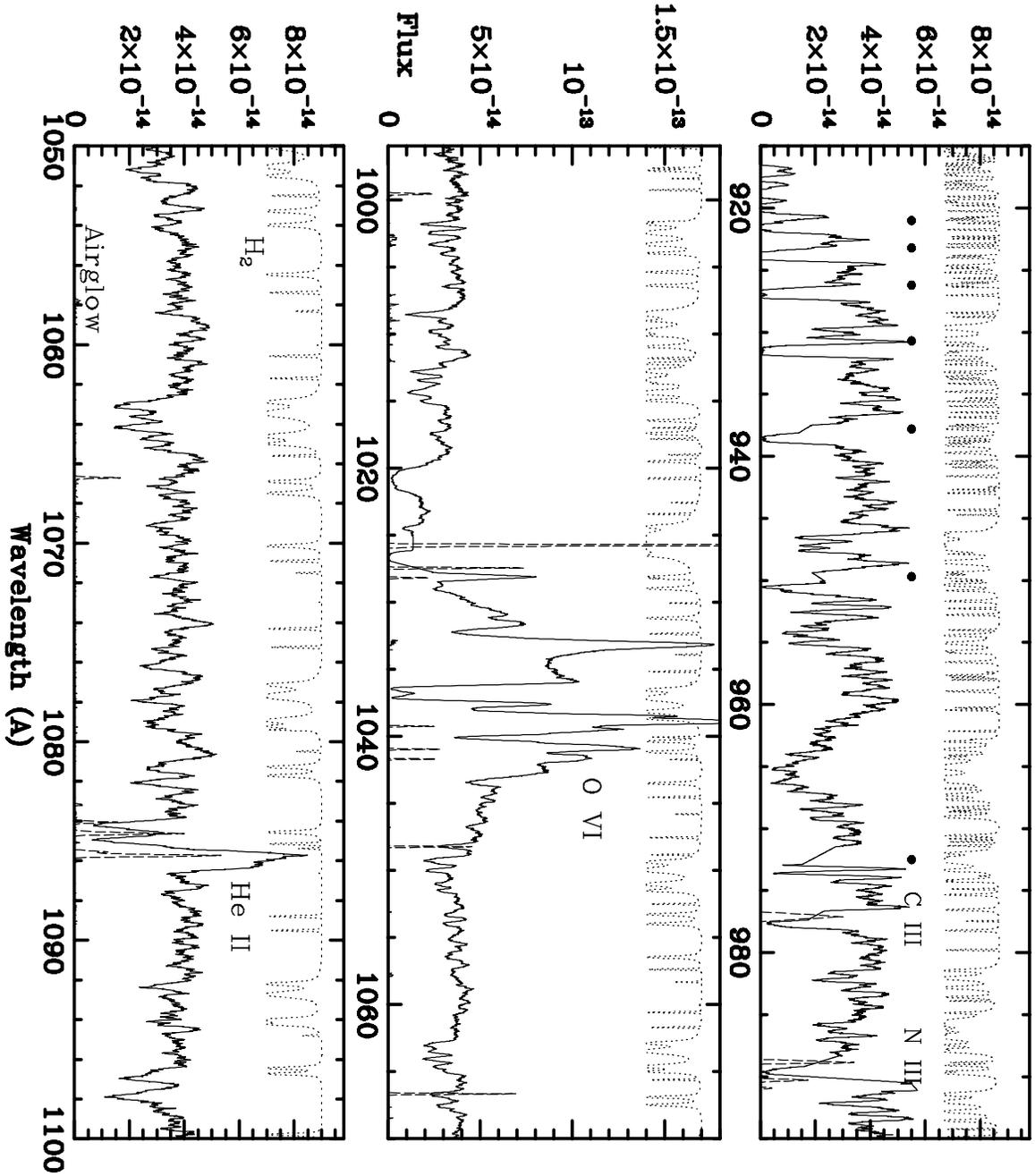}
\caption{The mean FUSE spectrum of RX~J0513.9$-$6951 (solid line).  The
dotted line at the top of each panel shows the positions and relative
strengths of H$_2$ absorbers, and the lower dashed line shows the
principal airglow features.  Fluxes are in erg cm$^{-2}$ s$^{-1}$ \AA$^{-1}$.
The dots in the upper panel show the rest wavelengths of Lyman lines.} 
\end{figure}

\begin{figure}
\plotone{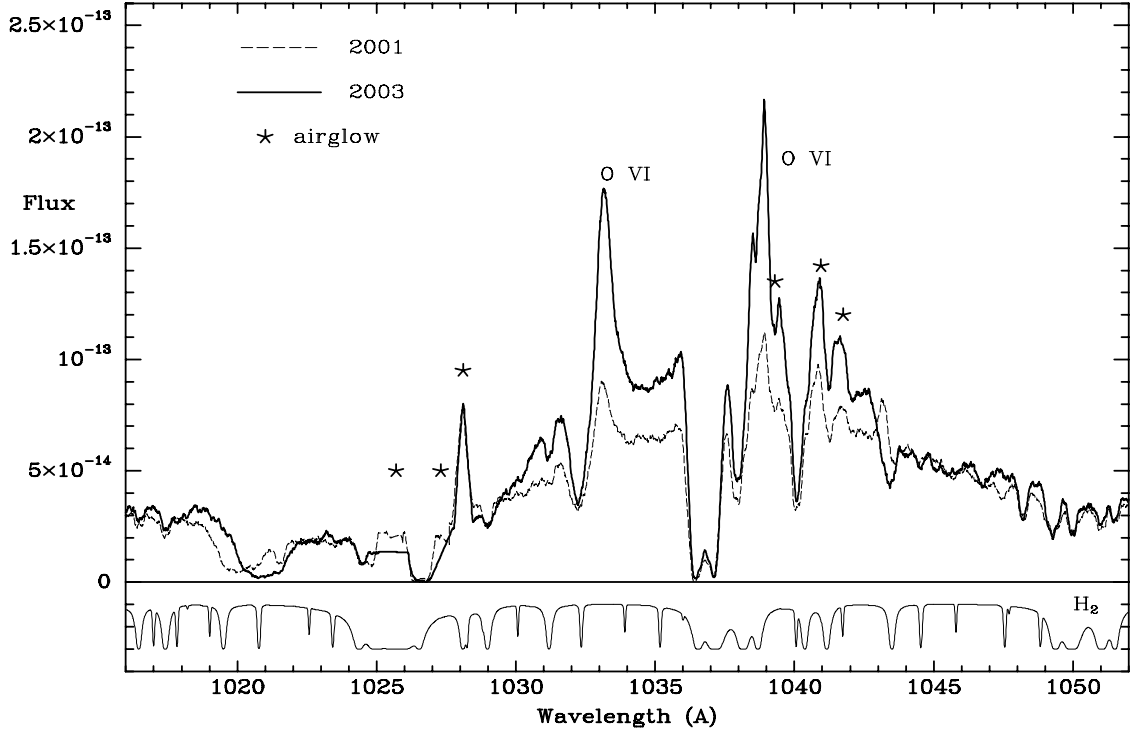}
\caption{Mean spectra in O~VI doublet region for the 2001 (dashed line)
and 2003 (solid line) epochs.  Fluxes are in erg cm$^{-2}$ s$^{-1}$
\AA$^{-1}$.  The zero point has been adjusted for the 2001 spectrum by
subtracting 1.3e-15, in order to match the continuum levels.  The overall
2003 profile is higher in the central region, while the airglow and
absorption features match well.  H$_2$ absorbers can be identified from
the model spectrum plotted in the lower panel.  The strongest airglow
lines (including Ly$\beta$ at 1026\AA) have been edited out manually, but
their positions are marked.  Note the change in the position of the
absorption feature near 1020-1\AA.} 
\end{figure}

\begin{figure}
\plotone{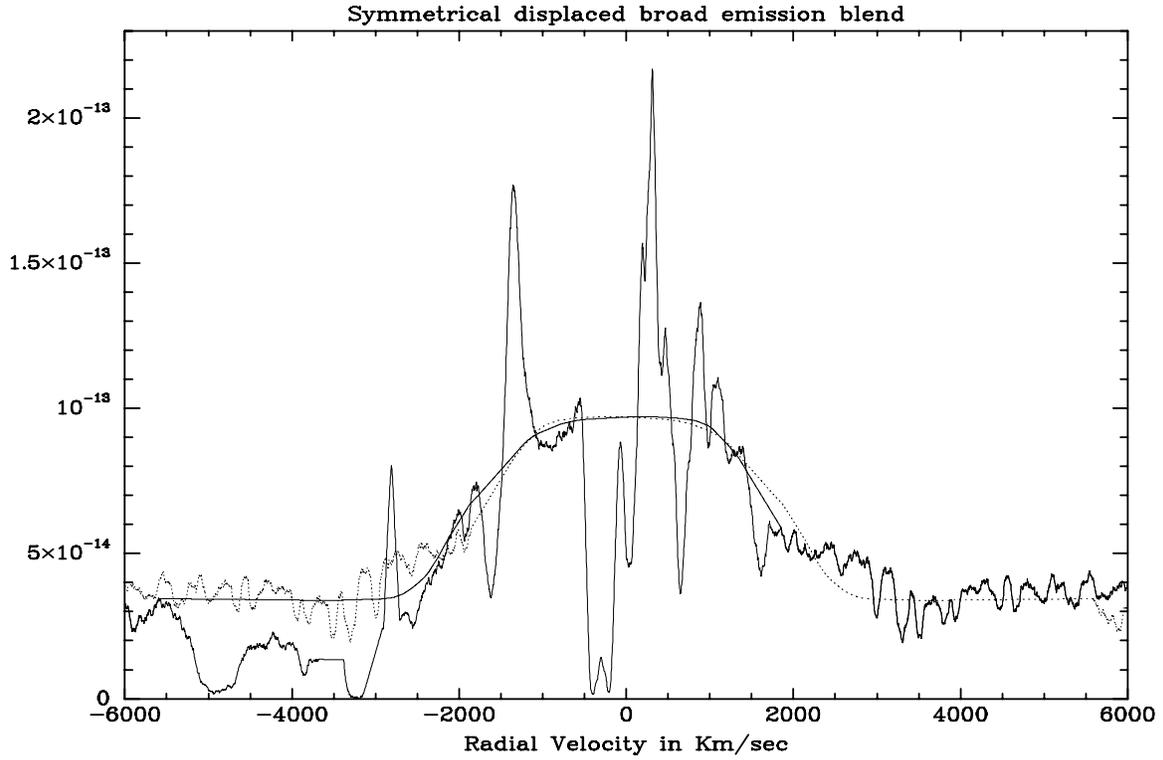}
\caption{Mean broad O~VI emission from the 2003 data, with the reflected
profile (shown as a dotted line) superimposed.  Fluxes are in erg cm$^{-2}$
s$^{-1}$ \AA$^{-1}$.  To create this profile, we have edited out the sharp
emissions and absorptions.  The center of the broad line is arbitarily set
to zero velocity.  The reflection wavelength is chosen to maximize the
broad feature symmetry.  Note that the center of the broad profile is
positively displaced with respect to the average position of the narrow
O~VI emission by $\sim$500 km s$^{-1}$, as discussed in the text.} 
\end{figure}
 
\begin{figure}
\plotone{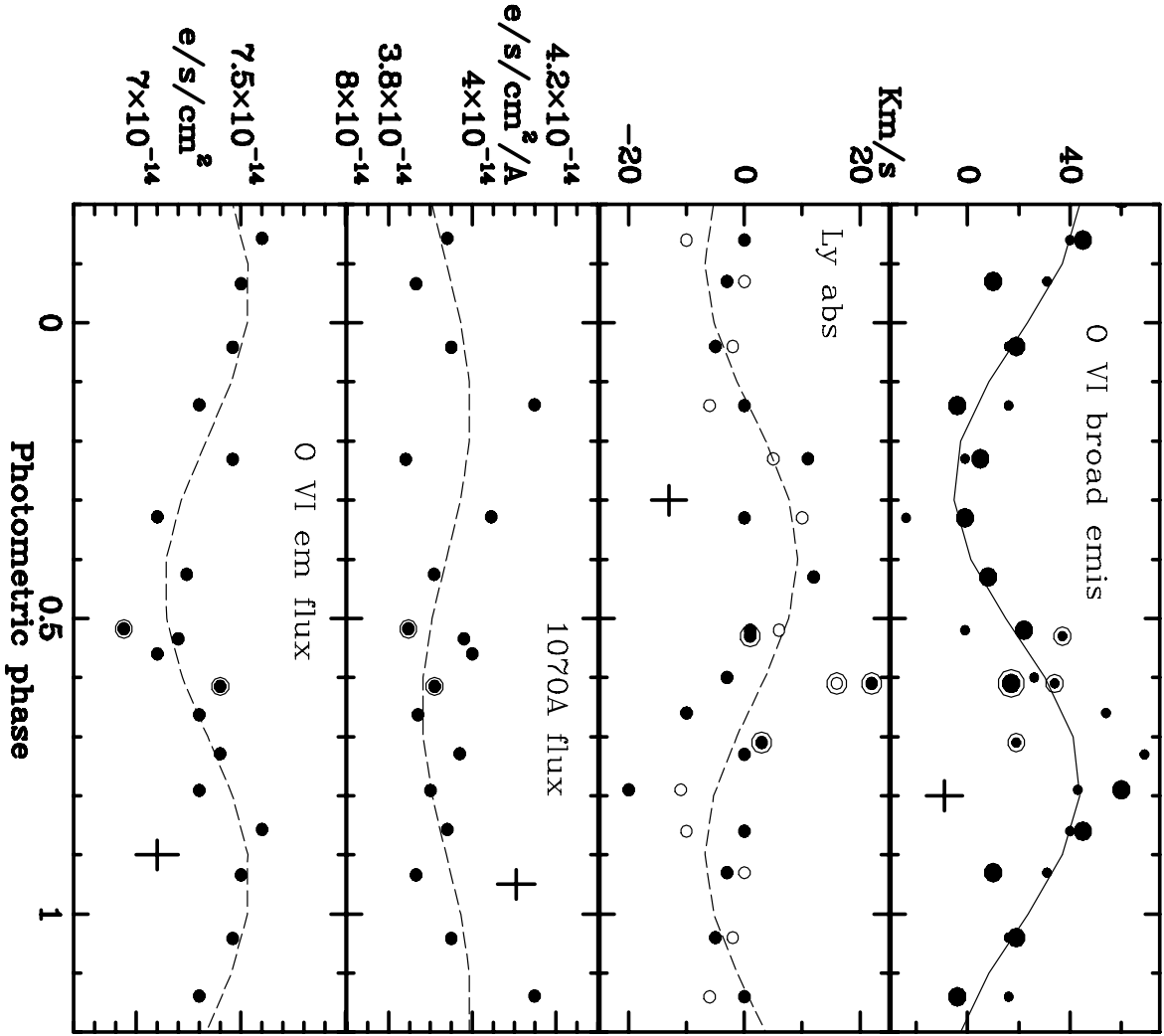}
\caption{Individual velocity measures with superimposed best-fit sine
curves.  The circled points are those from the final overlapping phases in 
the observation sequence. In a few cases, measures were not possible or
reliable due to data flaws, so not all quantities have 15 values plotted.
The crosses show the average formal 1$\sigma$ errors and phase range for 
a measure. (top) O~VI cross-correlation velocities from the 
LiF1A (large dots) and LiF2B (small dots).  (second) Lyman absorption 
velocities from cross-correlation of 915-940\AA\ region, from the SiC2A 
channel, with different symbols to denote the two measurement methods (see
text).  (third) Continuum flux from the LiF1A channel, in a region free of 
airglow and O~VI emission.  (bottom) Total flux from the broad O~VI emission, 
averaged from the LiF1A and LiF2B channels.} 
\end{figure}

\begin{figure}
\plotone{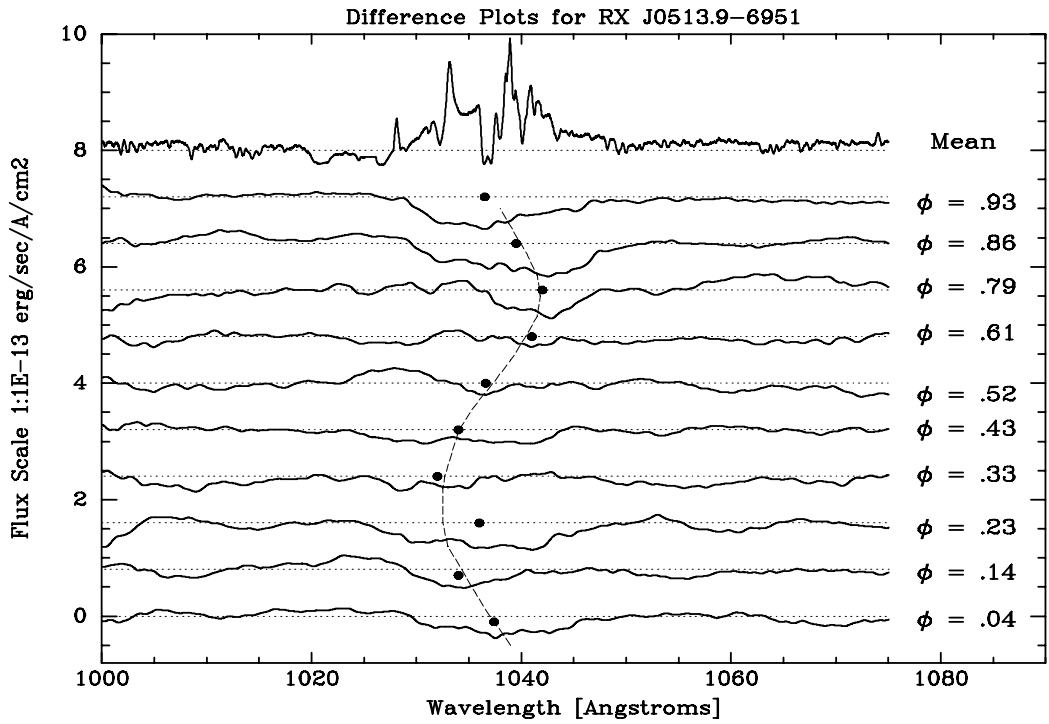}
\caption{Mean spectrum (top) and differences from the mean obtained by
subtracting the individual spectra.  The difference spectra are smoothed
over $\sim$5\AA\ and expanded by a factor 5.  The dots show the approximate
centroids of the broad absorption feature, and the dashed curve
illustrates a best-fit sine curve through them.} 
\end{figure}

\begin{figure}
\plotone{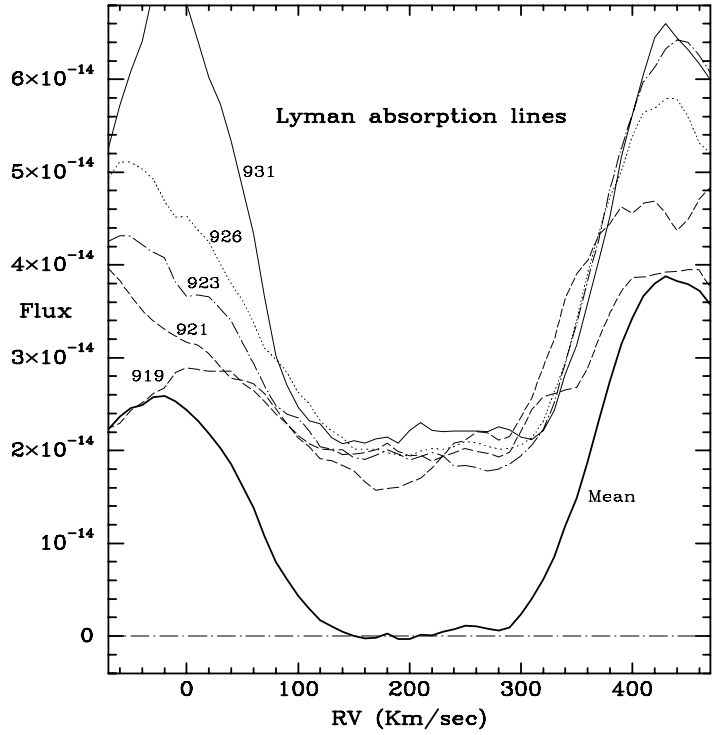}
\caption{Profiles of the Lyman absorption lines least contaminated by
airglow emission and H$_2$ absorption.  Fluxes are in erg cm$^{-2}$
s$^{-1}$ \AA$^{-1}$.  Each line is identified by its rest wavelength and
shifted vertically by 2.e-14 to provide separation from the mean profile
(shown as the heavy line).  Note the flat-bottomed saturated absorption
profiles and the large width in velocity.} 
\end{figure}

\begin{figure}
\plotone{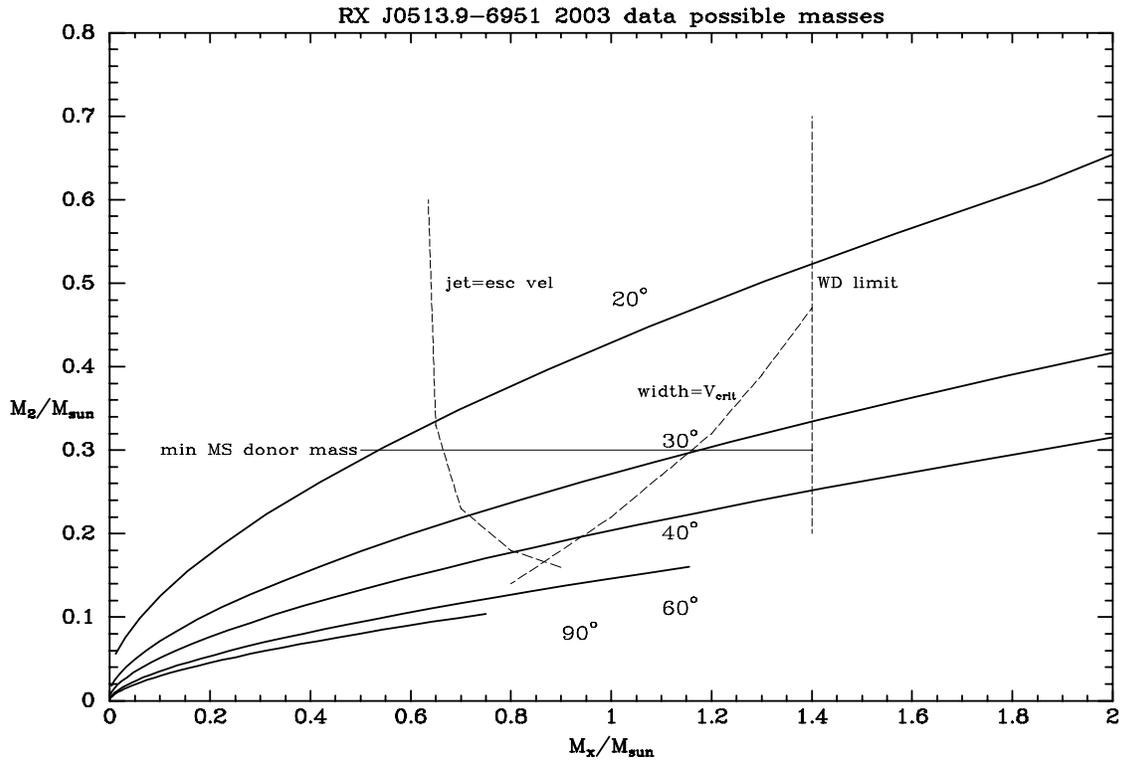}
\caption{Mass diagram resulting from adopting the overall 2003 broad O~VI
emission velocity curve and assuming it traces the orbital motion of the
compact star.  The sketched limits arise from other considerations of the
system, as described in the text.} 
\end{figure}

\end{document}